\documentclass[twocolumn]{revtex4}

\usepackage{graphicx}
\usepackage{bm}        
\usepackage{amsmath}
\usepackage{amssymb}   

\def\jcp#1#2#3{J.~Chem.~Phys.~{\bf #1},\ #2\ (#3)}

\def\pra#1#2#3{Phys.~Rev.~A~{\bf #1},\ #2\ (#3)}
\def\prl#1#2#3{Phys.~Rev.~Lett.~{\bf #1},\ #2\ (#3)}

\def\nat#1#2#3{Nature~{\bf #1},\ #2\ (#3)}

\def\njp#1#2#3{New~Journal~of~Physics~{\bf #1},\ #2\ (#3)}

\begin{document}

\title{Greatly enhanced absorption of non-resonant microwave fields by ultracold molecules near a Feshbach resonance
}

\author{S. V. Alyabyshev and R. V. Krems}
\affiliation{Department of Chemistry, University of British Columbia, Vancouver, V6T 1Z1, Canada}

\date{\today}

\begin{abstract}
We show that the probability of the collision-induced absorption of non-resonant microwave photons by a gas of ultracold molecules is dramatically enhanced near a Feshbach scattering resonance. This can be used for detecting Feshbach resonances of ultracold molecules by measuring the microwave field absorption and for tuning the elastic scattering cross sections of ultracold molecules by varying the frequency and intensity of the microwave field in a wide range of the field parameters.

\end{abstract}

\pacs{33.20.-t, 33.80.Ps}

\maketitle

The creation of ultracold atoms and molecules is projected to have an enormous impact on several different areas of physics research \cite{carr}, mainly because ultracold systems can be used for quantum simulation of complex phenomena in condensed-matter physics, precision measurements of extremely weak interactions and quantum computing. These applications require the ability to control ultracold gases externally. Ultracold atomic gases are usually controlled by varying an external magnetic field near a Feshbach scattering resonance \cite{tiesinga,ketterle,regal}. 
Magnetic Feshbach resonances indeed play a central role in the experimental research of ultracold atomic gases. 
They provide a mechanism to tune the scattering length of ultracold atoms. The sign and magnitude of the scattering length determines the dynamical properties of ultracold gases \cite{tiesinga,ketterle,regal}. 
Using magnetic fields for tuning ultracold gases, however, has two limitations: it is difficult to tune dc magnetic fields fast; and magnetic Feshbach resonances are usually very narrow, i.e. the scattering length of ultracold atoms can be tuned in a narrow range of magnetic fields. As a result, magnetic Feshbach resonances are often difficult to detect and the applications of magnetic field control of ultracold gases are limited to dynamical processes with the time scale of $>0.1$ ms \cite{bosenova}.

Many experiments have recently focused on the production of ultracold molecules in the ground internal energy state, successfully achieved by several research groups \cite{ni}.  Like ultracold atoms, ultracold molecules with non-zero magnetic moments can be controlled by means of magnetic Feshbach resonances \cite{yura}. However, the ro-vibrational structure of molecules allows for new possibilities of controlling molecular gases. In the present work, we show that magnetic Feshbach resonances of ultracold molecules can be modified by non-resonant microwave fields, leading to two important results: (i) the probability of collision-induced absorption of microwave photons is dramatically enhanced near a Feshbach resonance, which suggests a new method for detecting Feshbach resonances in collisions of molecules; and
(ii) the scattering length of ultracold molecules can be tuned by varying the frequency and intensity of microwave field, i.e. the scattering properties of ultracold molecules can be tuned in a wide range of microwave field intensities and frequencies. 

We consider an ensemble of polar diatomic molecules prepared in the absolute ground state and irradiated  with microwave field far detuned from molecular resonance.  In the absence of intermolecular interactions, the probability of microwave field absorption is very small. In the absence of microwave fields, molecules can undergo only elastic scattering. If molecules collide in the presence of non-resonant microwave field, the intermolecular interaction potential brings the rotational energy splitting of molecules in resonance with the microwave frequency. This leads to collision-induced absorption of microwave photons leading to the formation of a collision complex, in which one or both molecules are in a rotationally excited state. The molecules then undergo rotational relaxation, releasing energy, and the collision complex decays. This process is driven by the anisotropy of the intermolecular interaction potential \cite{mw}. The interaction with microwave fields thus induces inelastic decay channels, which suppress elastic scattering near Feshbach resonances \cite{hutson3}. We emphasize that this process is completely general and relies on two ingredients: (i) the presence of interparticle interactions that modifies the rotational energy splitting of the molecules; (ii) the anisotropy of the interparticle interactions that induces the rotational transitions in molecules leading to the break-up of the collision complex after the absorption of the photons. The strength of the microwave field can always be adjusted to compensate for the effect of the interaction anisotropy. To prove this, we consider two very different systems: NH($^3\Sigma$) molecules interacting with He atoms near a Feshbach resonance induced by a magnetic field and a pure gas of NH molecules near a Feshbach resonance induced by a dc electric field. The NH - He interactions are very weak. The NH - NH interactions are very strong.

  \begin{figure}
	\centering
	\includegraphics[width=0.5\textwidth, trim = 0 60 0 60]{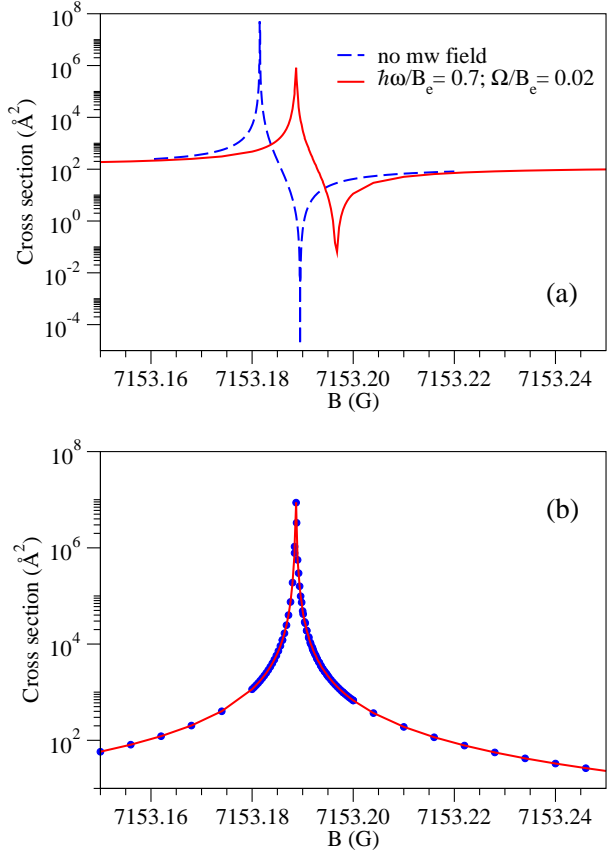}
	\renewcommand{\figurename}{Fig.}
	\caption{Cross sections for elastic collisions (panel a) and collisions accompanied by absorption of microwave photons (panel b) in  NH -- He scattering as functions of the magnetic field: broken line -- no microwave field; solid lines -- in the presence of a microwave field with $\Omega = 0.02 B_{\rm e}$ and $\hbar \omega = 0.7 B_{\rm e}$, where $B_e = 16.343$ cm$^{-1}$ is the rotational constants of NH. The line in panel (b) represents a sum of the transitions to all field-dressed states and the symbols represent the cross sections for the dominant single-photon transition $ | N=0, M_S = - 1, \bar{N} \rangle \rightarrow   | N=0, M_S = 1, \bar{N} - 1 \rangle$.
		} 	
\end{figure}
\noindent

First, we focus on low-energy collisions ($E_{\rm{kin}}=1~\mu$K) of NH molecules prepared in the rotationally ground state $N =0$  and the lowest-energy Zeeman level  $M_{S}=-1$ 
with $^{3}$He atoms near  a Feshbach resonance at $\sim$7150 G identified by Gonz{\'a}les-Mart{\'i}nez and Hutson \cite{hutson}.
The collision problem of molecules in the presence of a microwave field is best described using the field-dressed-state formalism described in detail in Ref. \cite{cohen-tannoudji}. The molecular system is assumed to be placed in a single-mode microwave cavity with parameters chosen to model a free-space maser beam \cite{cohen-tannoudji}. The Hamiltonian for the molecule in the presence of a dc magnetic and a single-mode microwave field is 
\begin{equation}
\hat{H}_{\rm as} = \hat{H}_{\rm mol} +\hat{H}_{\rm Z}+ \hat{H}_{\rm f} + \hat{H}_{\rm m, f},
\label{Has}
\end{equation} 
\noindent
where $\hat{H}_{\rm Z} = 2\mu_{B}{\bm{B}} \cdot \hat{S}$ describes the interaction of the total electron spin $S$ of the molecule with the dc magnetic field $\bm{B}$ and $\mu_{B}$ is the Bohr magneton.
The microwave field is described by $ \hat{H}_{\rm f} = \hbar\omega(\hat{a}\hat{a}^\dagger  - \bar{N}) $, where $\hat{a}^\dagger$ and $\hat{a}$ are the photon creation and annihilation operators and  $\bar{N}$ is a mean number of photons in the cavity \cite{mw2,cohen-tannoudji}.
The interaction of the linearly polarized microwave field of amplitude $E$ with the molecular dipole moment $d$ is given by $\hat{H}_{\rm m, f} =- \frac{\Omega}{2 \sqrt{\bar{N}}} (\hat{a} + \hat{a}^\dagger) \cos\phi$, where $\phi$ is the angle between the direction of the dipole moment and the polarization vector of the electric field and $\Omega = Ed$ describes the strength of the coupling of the molecule with the microwave field \cite{cohen-tannoudji}.
We assume that both the dc magnetic and microwave fields are directed along the quantization axis $z$.
The molecular Hamiltonian $\hat{H}_{\rm mol}$ describes the rotational and fine structure of the molecule \cite{hutson,potential}.
  We represent the total wave function of the system as a close coupling expansion in terms of the products of the eignefunctions 
of Hamiltonian (\ref{Has}) and the rotational wave functions of the collision complex (partial waves). The substitution of this expansion in the Schr\"{o}dinger equation leads to a system of coupled differential equations, which we solve numerically to obtain 
the probability of elastic scattering as well as collisions accompanied by absorption of photons and rotational relaxation \cite{mw,mw2}. 
Four rotational levels of the molecule, four photon number states and five partial waves were included in the basis set expansion, which leads to a system of 2880 equations for the total angular momentum projection equal to zero.  We used the potential energy surface for the He - NH collision complex calculated in Ref. \cite{potential}.


\begin{figure}
	\centering
	\includegraphics[width=0.5\textwidth, trim = 0 60 0 60]{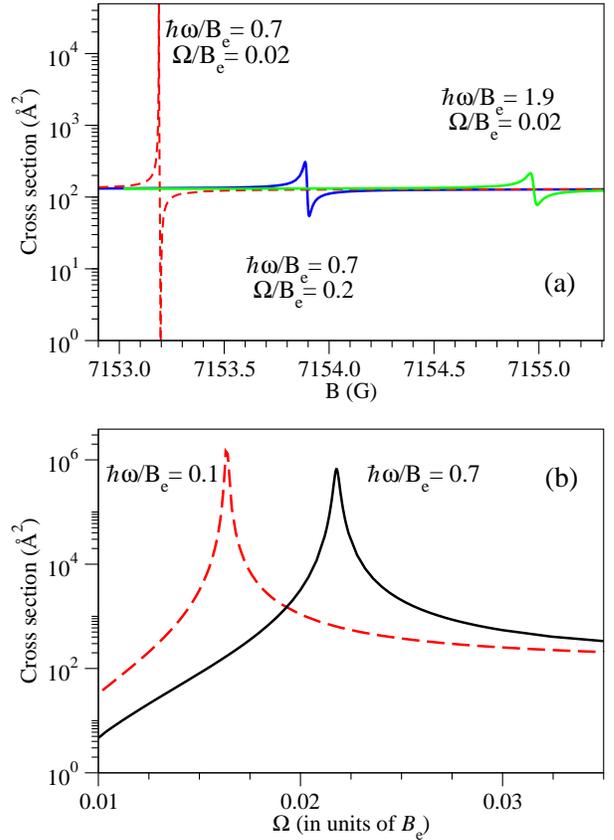}
	\renewcommand{\figurename}{Fig.}
	\caption{Cross sections for elastic collisions in  NH -- He scattering as functions of the magnetic field (panel a) and 
	 as functions of the microwave field strength (panel b) for different  parameters of the microwave field. The magnetic field in panel (b) is $B=7153.19$ G.
			} 	
\end{figure}
\noindent

\begin{figure}
	\centering
	\includegraphics[width=0.5\textwidth, trim = 0 60 0 60]{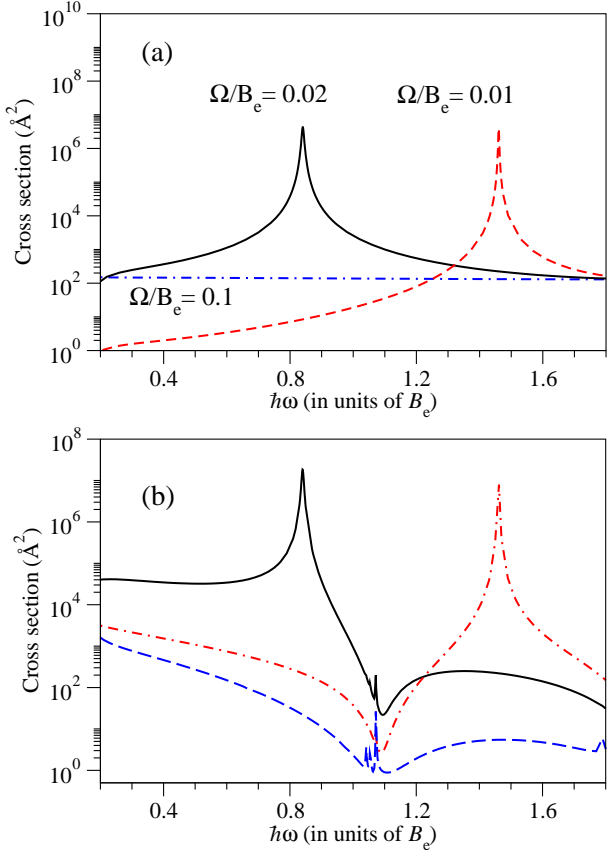}
	\renewcommand{\figurename}{Fig.}
	\caption{Cross sections for elastic collisions (panel a) and collisions accompanied by absorption of microwave photons (panel b) in  NH -- He scattering as functions of the microwave field frequency. The magnetic field is $B=7153.19$ G.  
	} 	
\end{figure}
\noindent

The eigenstates of Hamiltonian (\ref{Has}) are the field-dressed states of the molecule, i.e. they represent the combined molecule - field system. The solution of the coupled differential equations yields the scattering $S$-matrix that describes the probability amplitudes for transitions between the field-dressed states \cite{cohen-tannoudji,mw,mw2}. When the collision energy is smaller than the separation between the field-dressed states, only two types of collision processes can occur: elastic collisions and collisions accompanied by absorption of microwave photons. When molecules undergo elastic scattering they remain in the same field-dressed state. Inelastic relaxation brings the molecules from an initial field-dressed state to a field-dressed state of lower energy. Because all molecules are initially prepared in the rotationally ground state, the lower energy field-dressed states correspond to a lower mean number ($\bar{N}$) of photons in the cavity. Transitions to the lower energy field-dressed states must therefore be necessarily accompanied by absorption of photons.

Figure 1 shows the magnetic field dependence of the cross sections for NH - He collisions near the Feshbach resonance. The line in panel (b) is a sum of the cross sections for transitions to all field-dressed states except the initial state. The analysis of the state-resolved transitions shows that the cross section shown in panel (b) of Fig. 1 is dominated by a single transition $| \psi_{\rm i} \rangle \rightarrow | \psi_{\rm f} \rangle$, where $ | \psi_{\rm i} \rangle = a | N=0, M_S = - 1, \bar{N} \rangle  + b | ... \bar{N} ... \rangle$ and $ | \psi_{\rm f} \rangle = a | N=0, M_S = + 1, \bar{N} - 1 \rangle  + b | ... \bar{N} - 1 ... \rangle$. The ket $| ... \bar{N} ... \rangle$ denotes collectively the states mixed in the ground rotational state of the molecule due to molecule - field and fine-structure interactions. For weak, non-resonant microwave fields considered here, $a$ is very close to $1$. 
The cross section in panel (b) of Fig. 1 thus represents the probability of a process, in which the collision complex absorbs a photon, the molecule undergoes the spin flip and the rotational relaxation and the collision complex releases energy by decaying into the collision products.

Figure 1 demonstrates two important observations: (i) the position of the resonance in the presence of a microwave field is shifted; and (ii) the probability of collision-induced absorption of microwave photons (panel b) is dramatically enhanced near the Feshbach resonance. As predicted by Hutson \cite{hutson3}, the presence of strong inelastic transitions must suppress elastic scattering near a Feshbach resonance. This is indeed what we observe for collisions in microwave fields of higher intensity and lower detuning. The upper panel of Figure 2 shows that resonant enhancement of the elastic scattering cross section is suppressed, as the strength of the microwave field increases. 

\begin{figure}
	\centering
	\includegraphics[width=0.5\textwidth, trim = 0 60 0 60]{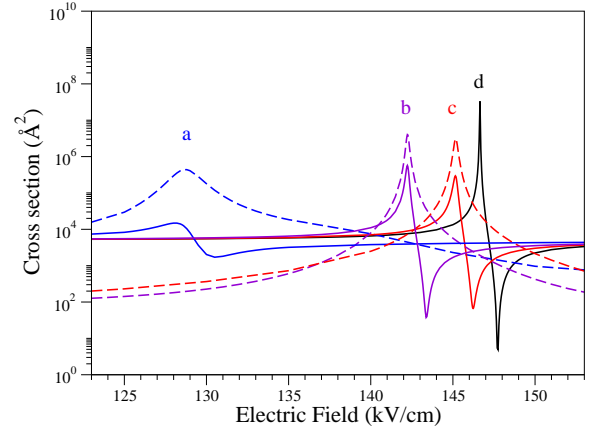}
	\renewcommand{\figurename}{Fig.}
	\caption{
	Cross sections for elastic collisions (solid lines) and collisions accompanied by absorption of microwave photons (dashed lines) in  NH -- NH scattering as functions of the electric field at a collision energy of 1 $\mu$K computed at  $\hbar \omega = 1.9 B_{\rm e}$ $\Omega = 0.04 B_{\rm e}$ (curve labeled $\it{a}$), $\hbar \omega = 1.9 B_{\rm e}$ $\Omega = 0.02 B_{\rm e}$ (curve labeled $\it{b}$),
	$\hbar \omega = 0.7 B_{\rm e}$ $\Omega = 0.04 B_{\rm e}$ (curve labeled $\it{c}$), and zero microwave field (curve labeled $\it{d}$).
		} 	
\end{figure}
\noindent

In order to elucidate the mechanism of microwave field modification of the magnetic Feshbach resonances, we use a model of the resonance in the presence of several decay channels \cite{hutson}.
The magnetic field dependence of the $S$-matrix elements is given by the following expression: 
\begin{equation}
{S}_{ i i'}(B) ={S}_{{\rm bg}, i i'} - \frac{{\rm i}g_{Bi}g_{Bi'}}{B - (B_{0}+\Delta)+{\rm i}\Gamma_B/2},
\label{Sii'}
\end{equation} 
\noindent
where ${S}_{bg,\rm i i'}$ is a slowly varying background scattering matrix element, $B_{0}$ is the resonance position in the absence of the microwave field, $\Delta$ is the shift of the position of the resonant level caused by the interaction with a microwave field, the energy dependent width of the resonance $\Gamma_{B}$ is given by the sum over the partial widths $\Gamma_{Bi}$ in accessible decay channels $i$: $\Gamma_{B}=\sum_{i}{\Gamma_{Bi}}$, and $g_{Bi}$ is a complex function so that $\Gamma_{Bi}=|g_{Bi}|^{2}$.
In the absence of the microwave field, the energy dependent width of the resonance is given by the single term $\Gamma_{B}=\Gamma_{B0} = 2k_{0}\gamma_{B0}$, where  $k_{0}$ is incoming wave vector and $\gamma_{B0}$ is energy-independent reduced width, which determines the variation of the scattering length as a function of the magnetic field. In the presence of a microwave field, $\Gamma_{B}$ includes contributions from decay channels $\Gamma_{B}^{\rm inel}=\sum_{i > 0}\Gamma_{Bi}$, which increase with increasing field strength.

The shift of the Feshbach resonances can be exploited for tuning the scattering properties of ultracold molecules with microwave fields. Figures 2 and 3 show the scattering cross sections near the Feshbach resonance as functions of the microwave field strength and frequency. The Feshbach resonance depicted in Figure 1 is very narrow, leading to the variation of the scattering cross section in a very small interval of magnetic fields $\sim 0.05$ G. While not impossible, it is technologically challenging to tune the dc magnetic field with such a high resolution. Figures 2 and 3 show that the same resonance gives rise to the variation of the scattering cross sections over wide ranges of the microwave field strength and frequencies.

In order to demonstrate that the results presented in Figures 1 - 3 are general, we consider a different system: NH - NH collisions near a Feshbach resonance induced by a dc electric field. For this calculation, we ignore the fine structure of the molecule, modify Hamiltonian (\ref{Has}) to include the interaction of molecules with a dc electric field, assume that the interacting molecules are distinguishable, use the basis of three rotational states for each molecule, two photon number states and four partial wave states and exploit the potential energy surface for the NH - NH complex in the quintet state calculated in Ref. \cite{janssen}. This calculation is not converged with respect to the basis set size and should be considered as a model used for illustrative purposes only. Figure 4 demonstrates that an electric-field-induced Feshbach resonance exhibits in the presence of a microwave field qualitatively the same suppression and shifts as the magnetic resonance examined in Figures 1 - 3. The absorption of microwave field at similar field parameters is much greater in NH - NH collisions than in NH - He collisions due to a much larger anisotropy of the NH - NH intermolecular potential.

In conclusion, we have demonstrated that the interaction of ultracold molecules with far-detuned, non-resonant microwave radiation is dramatically modified when molecules undergo resonant scattering. In particular, we showed that the probability of the collision-induced microwave field absorption is greatly enhanced in the vicinity of a Feshbach resonance. This suggests that measuring the microwave field absorption or loss of molecules as a function of microwave field parameters can be used as a method of detecting Feshbach resonances in an ultracold gas of molecules. Our calculations demonstrate that magnetic Feshbach resonances can be shifted in the presence of microwave fields by up to a few G. Feshbach resonances (including extremely narrow resonances) can therefore be located by varying the dc magnetic field with a step of a few G and scanning the intensity of a non-resonant microwave field at a fixed magnetic field.  

The interaction with non-resonant microwave fields induces inelastic losses of molecules due to collision-induced absorption of microwave photons. 
This leads to suppression of Feshbach resonances and allows for tuning Feshbach resonances by varying both the intensity and frequency of the microwave field. 
Laser fields can be tuned much faster than dc magnetic fields. For example, the collapse of a BEC can be induced in experiments with ultracold atoms by varying the magnetic field near a Feshbach resonance on the time scale of 0.1 ms \cite{bosenova}. The variation of the microwave fields depicted in Figures 2 and 3 can be achieved on the time scale of nanoseconds. This can be used for new studies of BEC dynamics with instabilities induced on a much shorter time scale or BEC dynamics with instabilities oscillating on the time scale of intermolecular interactions. Finally, because the absorption of microwave field is enhanced near a Feshbach resonance, our results suggest that a combination of a Feshbach resonance and microwave fields can be used for photoassociation of atom - molecule or molecule - molecule collision complexes to produce ultracold trimers and tetramers. This technique would be analogous to the technique of Feshbach-enhanced photoassociation of ultracold atoms \cite{cote}. 

We thank Timur Tscherbul for useful discussions and for providing the Fortran code generating the PES for the NH - NH complex. 
This work was supported by NSERC of Canada and the Canadian Foundation for Innovation.

\end{document}